\documentclass[journal,twoside]{IEEEtran}
\usepackage{cite}
\usepackage{amsmath,amssymb,amsfonts}
\usepackage{graphicx}
\usepackage{textcomp,nicefrac}

\def\BibTeX{{\rm B\kern-.05em{\sc i\kern-.025em b}\kern-.08em
T\kern-.1667em\lower.7ex\hbox{E}\kern-.125emX}}
\begin{document}
\title{Design of the Readout Electronics for the TRIDENT Pathfinder Experiment}
\author{M. X. Wang, G. H. Gong, P. Miao, Z. Y. Sun, J. N. Tang, W. H. Wu, and D. L. Xu 
\thanks{Manuscript received XX XX, 2022; revised XX XX, 2022 and
	XX XX, 2022; accepted XX XX, 2022. Date of publication XX XX,
	2022; date of current version XX XX, 2022. This work was supported in
	part by the Shanghai Pujiang Program (No. 20PJ1409300); in part by Shanghai Pilot Program for Basic Research - Shanghai Jiao Tong University (No. 22Z511203770); in part by the Ministry of Science and Technology of China
	(No. 2022YFA1605500); and in part by Office of Science and Technology, Shanghai Municipal Government (No. 22JC1410100). The authors would like to thank the sponsorship from Yangyang Development Fund.}
\thanks{M. X. Wang, Z. Y. Sun, J. N. Tang, and W. H. Wu are with the School of Physics and Astronomy, Shanghai Jiao Tong University, Shanghai 200240, China.}
\thanks{G. H. Gong is with the Department of Engineering Physics, Tsinghua University, Beijing 100084, China.}
\thanks{P. Miao is with the Department of Modern Physics, University of Science and Technology of China, and also with the State Key Laboratory of Particle Detection and Electronics, USTC, Hefei 230026, China.}
\thanks{D. L. Xu is with the Tsung-Dao Lee Institute, Shanghai Jiao Tong University, Shanghai 201210, China, and also with the School of Physics and Astronomy, Shanghai Jiao Tong University, Shanghai 200240, China.}
\thanks{Corresponding Author: W. H. Wu (e-mail: wuweihao@sjtu.edu.cn).}}

\maketitle

\begin{abstract}
The tRopIcal DEep-sea Neutrino Telescope (TRIDENT) is a future large-scale next-generation neutrino telescope. In September 2021, the TRIDENT pathfinder experiment (TRIDENT EXplorer, T-REX for short) completed \emph{in-situ} measurements of deep-sea water properties in the South China Sea. The T-REX apparatus integrates two independent and complementary systems, a photomultiplier tube (PMT) and a camera system, to measure the optical and radioactive properties of the deep-sea water. One light emitter module and two light receiver modules were deployed, which were synchronized by using White Rabbit (WR) technology. The light emitter module generates nanosecond-width LED pulses, while the light receiver module hosts three PMTs and a camera to detect photons. The submerged apparatus and the data acquisition system (DAQ) perform real-time command and data transmission. We report the design and performance of the readout electronics for T-REX, including hardware modules, firmware design for digital signal processing, and host-computer software.
\end{abstract}

\begin{IEEEkeywords}
Readout electronics, data acquisition, real time systems, optical measurements, marine technology.
\end{IEEEkeywords}

\section{Introduction}
\label{sec:introduction}
\IEEEPARstart{H}{igh-energy} astrophysical neutrinos carry unique information about the extreme environment of the universe. Since neutrinos rarely interact with matter, a detector with the size of cubic kilometers housed in water or ice is required to measure the astrophysical neutrino flux. Neutrino astronomy is booming since the IceCube Neutrino Observatory\cite{icecube1} first discovered a diffuse astrophysical neutrino flux\cite{icecube2}. While IceCube has put strong constraints on the intensity of neutrino sources, discovering weak neutrino sources calls for the next-generation neutrino telescope with better angular resolution\cite{Ackermann:2019ows}. At present, several next-generation neutrino telescopes are under construction or proposed worldwide, including IceCube-Gen2\cite{icecubegen2}, KM3NeT\cite{km3net}, Baikal-GVD\cite{baikal}, and P-ONE\cite{pone}.

The tRopIcal DEep-sea Neutrino Telescope (TRIDENT)\cite{TRIDENTphy} is a proposed future next-generation neutrino telescope which covers $\sim$8 km$^{3}$ volume of seawater and will be constructed on the ocean floor of the South China Sea. It will become an important part of the global network of neutrino observatories to provide precise measurements of cosmic neutrinos. TRIDENT comprises $\sim$1000 strings. Each string consists of a base, vertical electro-optical cables, and hybrid digital optical modules (hDOMs). Each hDOM contains multiple photomultiplier tubes (PMTs) and silicon photomultipliers (SiPMs) to detect the Cherenkov light induced by relativistic particles emerging from neutrino interactions \cite{hdom}.

A sea area eligible for the construction of neutrino telescopes needs to meet the conditions of having a large flat seabed topography and seawater with the required level of transparency. As the first site selection mission for TRIDENT, the TRIDENT pathfinder experiment (TRIDENT EXplorer, T-REX for short) aims to measure the optical properties of seawater and the ambient background light of the site \emph{in-situ} for the future neutrino telescope, as well as measure the seabed topography and oceanographic conditions of the site.

\begin{figure}[htbp]
	\centerline{\includegraphics[width=3.5in]{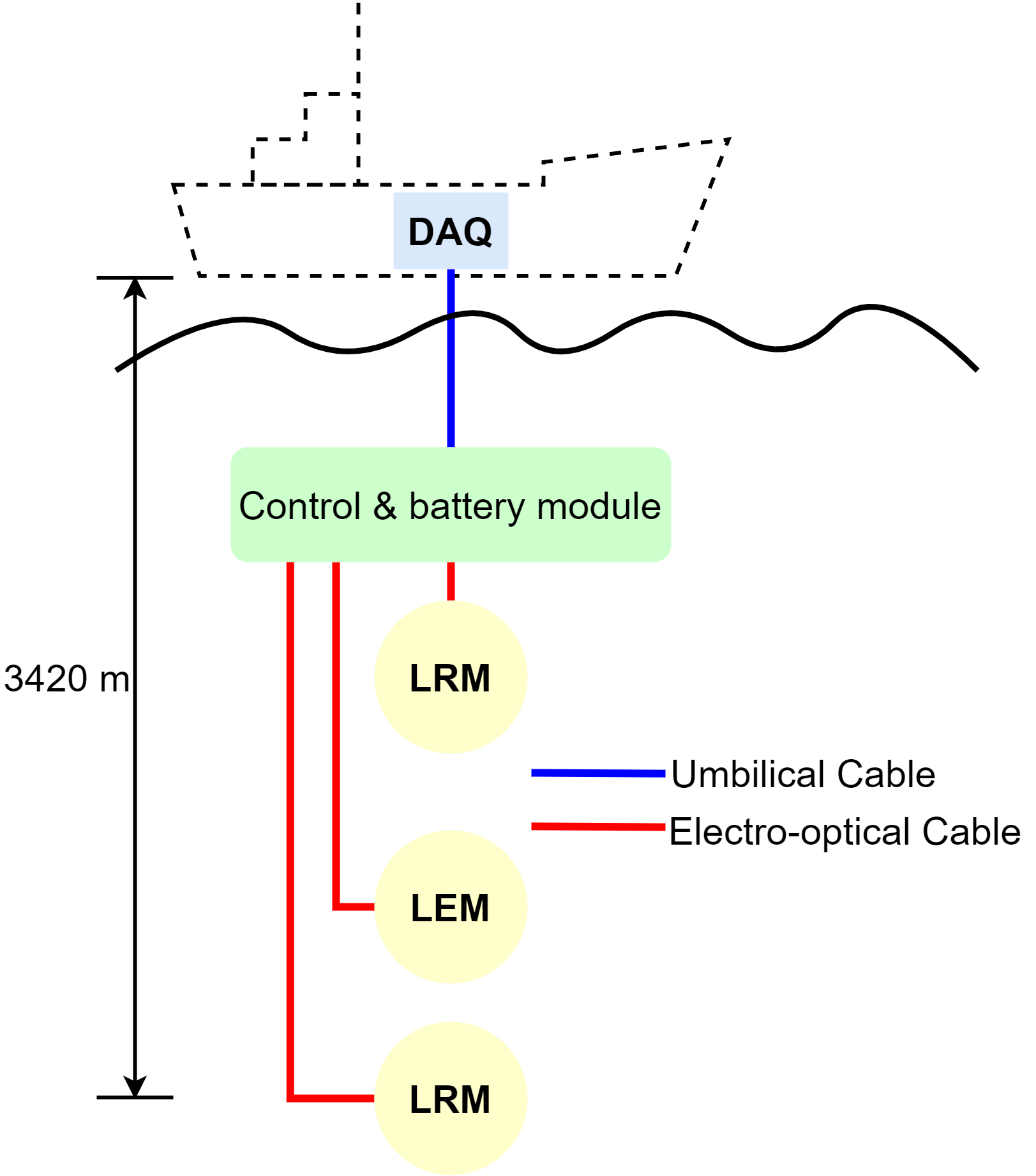}}
	\caption{A schematic plot of the T-REX detection system (not to scale). The dashed box on the top is the DAQ on the research vessel. The underwater apparatus is mechanically connected in series. The umbilical cable is the link between the DAQ and the control and battery module. Each module is connected to the control and battery module by a corresponding electro-optical cable.}
	\label{fig:e_o_d}
\end{figure}

T-REX was carried out in September 2021. The T-REX apparatus was deployed from a research vessel to a depth of 3420 m by an umbilical cable. As shown in \figurename~\ref{fig:e_o_d}, the umbilical cable connected the data acquisition system (DAQ) with the control and battery module. The DAQ, which was deployed in the laboratory on the research vessel during the experiment, sent commands and processed data. One of the four fibers in the umbilical cable was used for the control board communication in the control and battery module, and the remaining fibers were for other modules. In the vertical direction, the light emitter module (LEM) was located between two light receiver modules (LRMs), and the distances relative to the LRMs were 21.73$\pm$0.02 m and 41.79$\pm$0.04 m, respectively\cite{TRIDENTphy}. Two hemispheres of LEM could illuminate both LRMs simultaneously. The photosensitive area of the LRM faced the LEM. Each module was connected to the control and battery module by a corresponding electro-optical cable.

This paper starts with a description of the requirements for the detector, followed by a detailed description of the design of the readout electronics. After introducing results from the laboratory tests, a conclusion is given.

\section{Requirements from the detector}
\label{sec:detector}
The absorption and scattering of Cherenkov photons in seawater strongly impact the reconstruction of physical events. The most important optical properties measured in this experiment are therefore the absorption length and scattering length of the deep-sea water at the wavelength range between 400 nm and 550 nm, to which Cherenkov light belongs. For this purpose, light-emitting diodes (LEDs) of different wavelengths (405 nm, 450 nm, 460 nm, 525 nm) are used\cite{LEM}. Two independent schemes, pulsing and steady modes, are used in this experiment.

For pulsing mode, both the upper and lower hemispheres of the LEM simultaneously flash with a light pulse of several nanoseconds width. The single-photon response of the PMT (XP72B22, Hainan Zhanchuang Photonics Technology Co., Ltd) has a high temporal precision, which is used to detect pulsing light. The intensity of the light pulse emitted by the LEM needs to lead to a signal amplitude at the level of a single photoelectron (SPE) in the LRM at $\sim$40 m. Since the typical PMT SPE signal has a 4 mV amplitude and 4 ns rise time, a low-noise preamplifier and a 12-bit 250 MSPS analog-to-digital converter (ADC) are used to obtain PMT SPE waveforms. 
Some of those photons detected by the LRM have been scattered multiple times by seawater. Hence the tail of the photon arrival time distribution in the LRM, which can extend to $\sim$100 ns\cite{TRIDENTsim}, represents the scattering effect of the seawater.
The experiment samples six PMTs simultaneously in 1-microsecond windows with a global trigger mode. To collect sufficient statistics in about two hours, the pulsers are operated at a frequency of 10 kHz. Accordingly, the readout system must operate stably at 300 Mbps. 

For steady mode, cameras in the LRMs record images of the continuously glowing LEM to measure the angular distribution of luminous brightness, which is affected by scattering and absorption\cite{TRIDENTsim, camera}. Camera images are uploaded to the DAQ on the research vessel for real-time monitoring and saved to an SD card as a backup simultaneously.

As mentioned in Section~\ref{sec:introduction}, one of the important criteria for site selection is the level of ambient background light. The ambient background mainly includes bioluminescence produced by deep-sea organisms, light from the nuclear decay of $^{40}$K dissolved in seawater, and radioactive decay in the glass sphere. 
The experiment samples all PMTs in self-trigger mode to measure the ambient background and requires that the normal operation of the acquisition channel is immune to the paralysis of any other channel brought on by the abrupt increase in event rate.

\begin{figure}[htbp]
	\centerline{\includegraphics[width=3.5in]{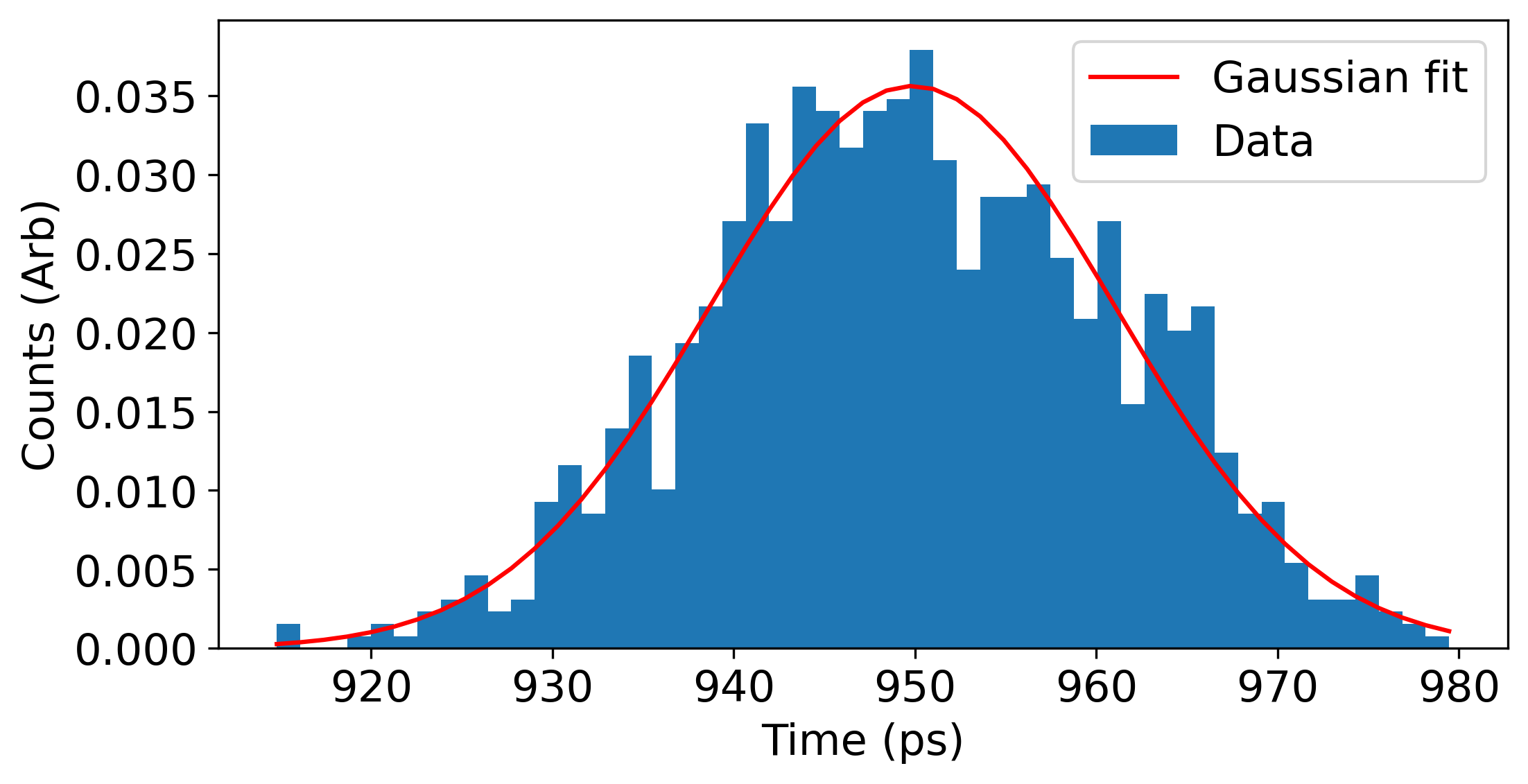}}
	\caption{The histogram illustrates the measured skew between the PPS of a CLB and the PPS of a WR switch. The red line is a Gaussian fit to the data with 11-ps standard deviation.}
	\label{fig:wrrms}
\end{figure}

The experiment aims to measure the absolute time of arrival of each photon from the pulsing LEM. The clock synchronization system uses the same White Rabbit (WR) technique as the Large High Altitude Air Shower Observatory (LHAASO) project \cite{wrLHAASO}. The timing distribution of the WR system has sub-nanosecond accuracy and tens of picoseconds precision \cite{wrper}. The WR node \cite{wrnode} in each module communicates with the WR switch \cite{wrswitch} in the laboratory of the research vessel via optical fibers.
The pulse per second (PPS) and 62.5 MHz clock output by each WR node are highly synchronized. The skew between the PPS of a central logic board (CLB) and the PPS of a WR switch has been examined in the laboratory to qualify the stability of the synchronization. As shown in \figurename~\ref{fig:wrrms}, the skew has a Gaussian distribution with an 11-ps standard deviation. 

A wavelength division multiplexer (WDM) system is used for the communication between DAQ and modules. Each LEM and LRM occupies one fiber with a four-channel compact coarse wavelength division multiplexer (4cCCWDM) inside. In addition, multiple adapters and connectors lead to large optical loss, which indicates that the light attenuation is more than 20 dB, and small form-factor pluggables (SFPs) for long-distance communication are required. 

To meet the requirements of T-REX, the design of the readout electronics focuses on the following features:
\begin{enumerate}
	\item[$\bullet$] The hybrid detector integrates two isolated and complementary systems, a PMT and a camera system, to measure the optical and radioactive properties of seawater. These two distinct measurements are completed alternately within a finite time since the two systems operate under different light modes.
	\item[$\bullet$] The WR network synchronizes onboard clocks in each module with sub-nanosecond precision. This technique will also be applied to TRIDENT\cite{TRIDENTphy}.
	\item[$\bullet$] A tunable pulsed light source can emit nanosecond-wide pulses of light and illuminate about 40 meters away.
	\item[$\bullet$] During the experiment, the apparatus and research vessel are stably connected through an umbilical cable. The condition of the entire detector is monitored and the operating configuration of the detector can be adjusted in real time with the help of the slow control system, which increases the flexibility of the measurement strategy.
\end{enumerate}

The following sections present the electronics design that satisfies the requirements of the experiment.

\section{Design of the readout electronics}
\label{sec:DOMele} 
\begin{figure}[htbp]
	\centerline{\includegraphics[width=3.5in]{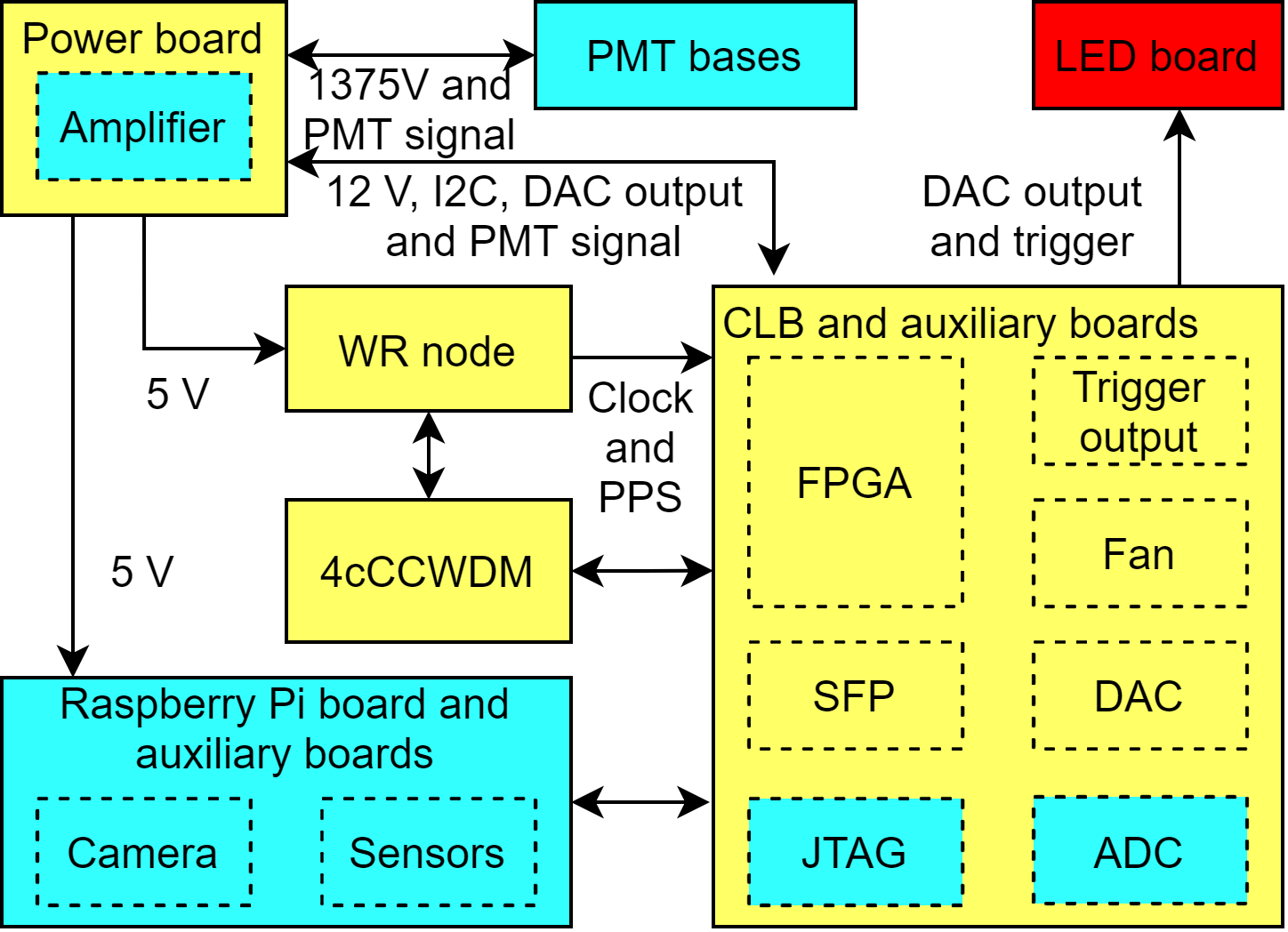}}
	\caption{Block diagram of the electronic boards in a module (LRM or LEM) and their interconnections. Yellow sections are present in each module. Blue sections are only in the LRM, with the red section only in the LEM. The power board receives the 14.8 V input, generates all the voltages needed by the rest of the electronic boards, and transfers the amplified PMT signals to the ADC mezzanine board. The CLB, the main electronic board, includes an FPGA that runs the data acquisition firmware.}
	\label{fig:domhd}
\end{figure}

As mentioned in Section~\ref{sec:introduction}, there are two types of modules: LEM and LRM. The block diagram of the electronics system in the module is in \figurename~\ref{fig:domhd}, which visualizes the design concept of modules with some differences between the LRM and the LEM. 

\begin{figure}[htbp]
	\centerline{\includegraphics[width=3.5in]{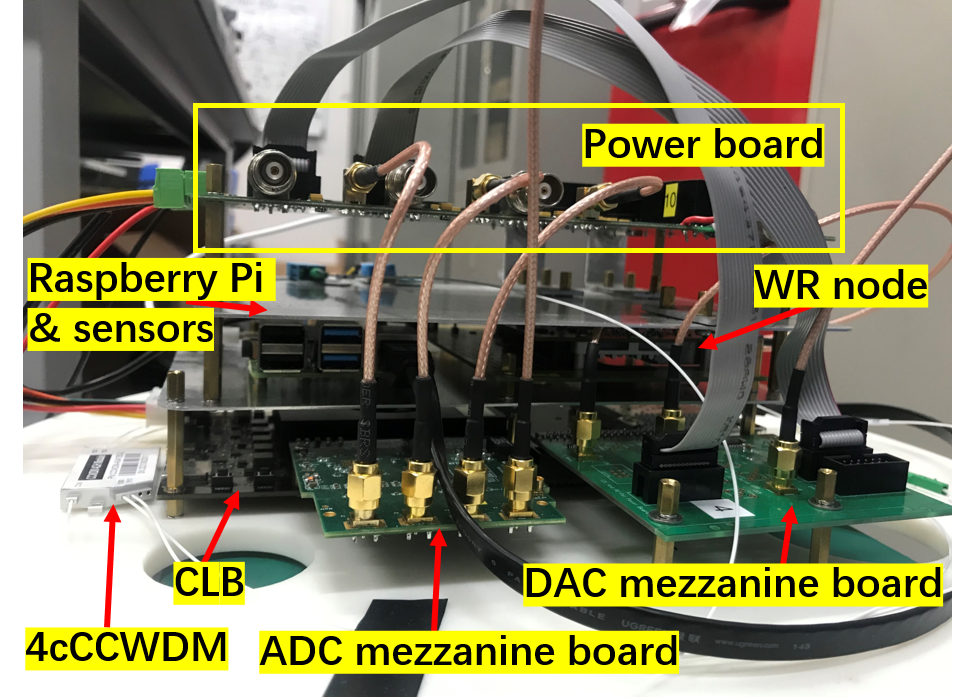}}
	\caption{Horizontal view of the electronics system in the LRM.}
	\label{fig:abd}
\end{figure}

One hemisphere of the LRM contains three PMTs in different orientations with a camera in the middle. In the other hemisphere, the electronic boards are fixed in a three-layer structure, shown in \figurename~\ref{fig:abd}. The CLB (ALINX AX7325) and a 4cCCWDM are hosted at the bottom 3D-printed layer.
On the second layer, the Raspberry Pi and WR node are fixed with copper pillars on an aluminum plate. In order to protect the sensitive analog signal input from the high-frequency noise generated by DC/DC converters in the LRM, the power board (PB) is installed on top of the bracket with an aluminum plate and copper pillars.

All aluminum plates and copper pillars are connected for heat dissipation and electromagnetic shielding. The controlled fan mounted on the CLB blows gas to the field programmable gate array (FPGA) and adjacent power supply modules, which also enhances the gas flow in the sphere. Laboratory tests show that the FPGA temperature is reduced from $70\,^{\circ}\mathrm{C}$ to $43\,^{\circ}\mathrm{C}$ with the help of the running fan. When the ambient temperature near the seabed is $\sim2\,^{\circ}\mathrm{C}$, the measured temperature of the FPGA is $30\,^{\circ}\mathrm{C}$.

\begin{figure}[htbp]
	\centerline{\includegraphics[width=3.5in]{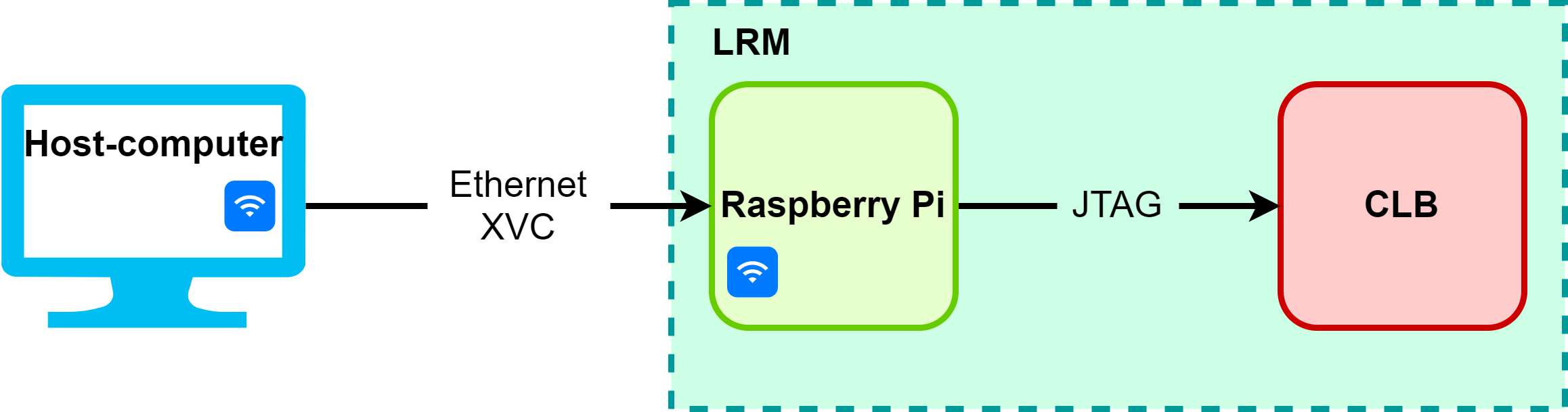}}
	\caption{Block diagram of remote access to the FPGA on CLB. The Xilinx virtual cable protocol supports the host-computer to configure the FPGA through the Raspberry Pi. Before deploying the apparatus, the host-computer and Raspberry Pi can use wireless communication.}
	\label{fig:xvc}
\end{figure}

With the Raspberry Pi WiFi module, the FPGA firmware can be updated wirelessly, without opening the capsuled module. As illustrated in \figurename~\ref{fig:xvc}, the Vivado design tools can communicate the same JTAG commands over a TCP/IP connection to the Raspberry Pi implementing the Xilinx virtual cable (XVC) protocol \cite{xvc}.  

\subsection{Digitization of PMT Signals}
\label{subsec:PMT}
\begin{figure}[htbp]
	\centerline{\includegraphics[width=3.5in]{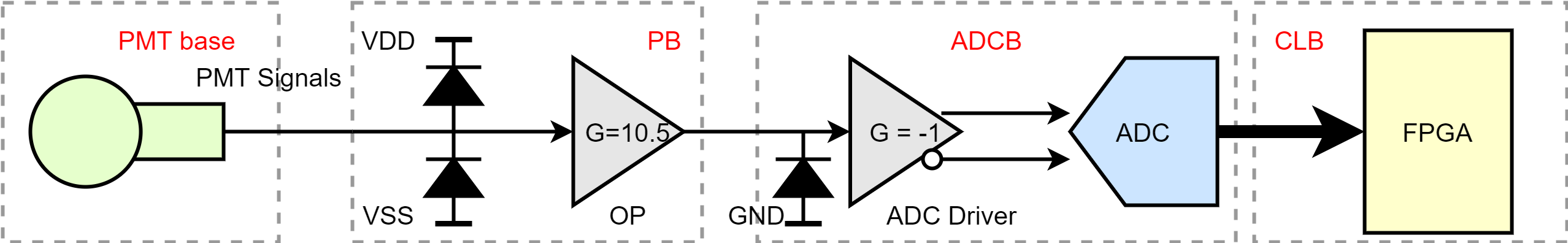}}
	\caption{Block diagram of the analog front-end circuit. The PMT signal is amplified by the preamplifier on the PB and then converted into a differential signal on the ADC mezzanine board (ADCB). The differential signal is digitized by a 250 MSPS ADC and buffered in the FPGA.}
	\label{fig:preamp}
\end{figure}

In T-REX, the PMT SPE signals are used to analyze the distribution of the arrival time of photons. As the typical SPE signal has an amplitude of 4 mV, a low-noise high-bandwidth preamplifier is adopted. \figurename~\ref{fig:preamp} shows the block diagram of the analog front-end circuit. 

\begin{figure}[htbp]
	\centerline{\includegraphics[width=3.5in]{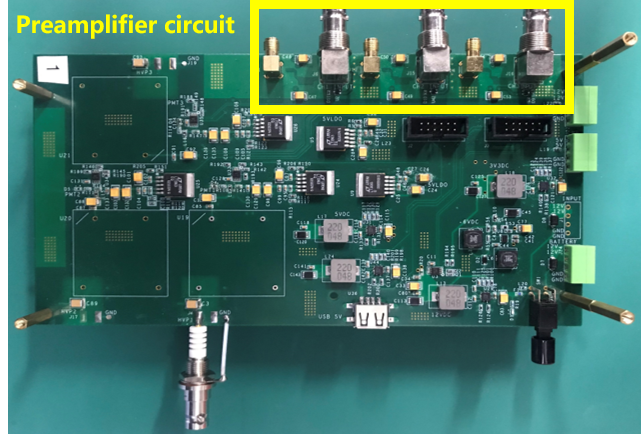}}
	\caption{The photograph of the power board (the high voltage modules are uninstalled). There are three preamplifier circuits in the yellow box.}
	\label{fig:pb}
\end{figure}

The PMT signal is output through AC coupling since the positive high voltage is applied to the anode of the PMT. Notably, the first stage preamplifiers (OP, ADI AD8009) with $\times10.5$ gain and 300 MHz bandwidth (-3 dB) are located on the PB (see \figurename~\ref{fig:pb}), based on the suppression of electromagnetic interference and the convenience of integration.

The second stage of amplification on the ADC mezzanine board (ADCB) comprises an inverter (ADI AD8009) and an ADC driver (ADI AD8138), which can further magnify the signal according to experimental requirements, but the default is unity gain.

The front-end circuit comprises transient-voltage-suppression diodes (TI ESD321DPYR) and a high-speed switching diode (Nexperia BAV99W) to clamp the large transient current which stems from the electrostatic discharge (ESD) and possible overvoltage spark from high voltage components in the PMT base. Thanks to low diode capacitance, PMT signals are scarcely affected by the protection circuit. 

\begin{figure}[htbp]
	\centerline{\includegraphics[width=3.5in]{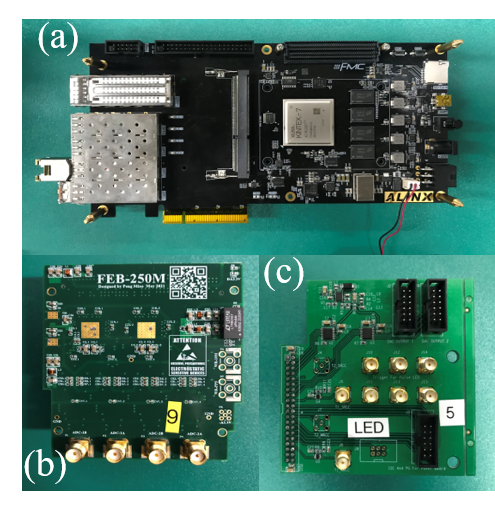}}
	\caption{The photographs of the CLB (the fan above the FPGA in the picture has been removed) (a), the ADCB (two ADC chips on the back of the middle pad) (b) and the digital-to-analog mezzanine board (DACB) (c). }
	\label{fig:CLBbs}
\end{figure}

A custom-made four-channel ADCB is used for PMT waveform digitization (see \figurename~\ref{fig:CLBbs}(b)). Two ADI AD9613 chips on the board with a sampling rate of 250 MSPS and a 12-bit resolution simultaneously sample three channels of PMT signals and one channel of the global trigger as a verification. The digital data are directly transmitted to the FPGA (Xilinx Kintex-7 FPGA, XC7K325TFFG900) for event building.

\subsection{Power supply in modules}
\label{subsec:powerindom}
\begin{figure}[htbp]
	\centerline{\includegraphics[width=3.5in]{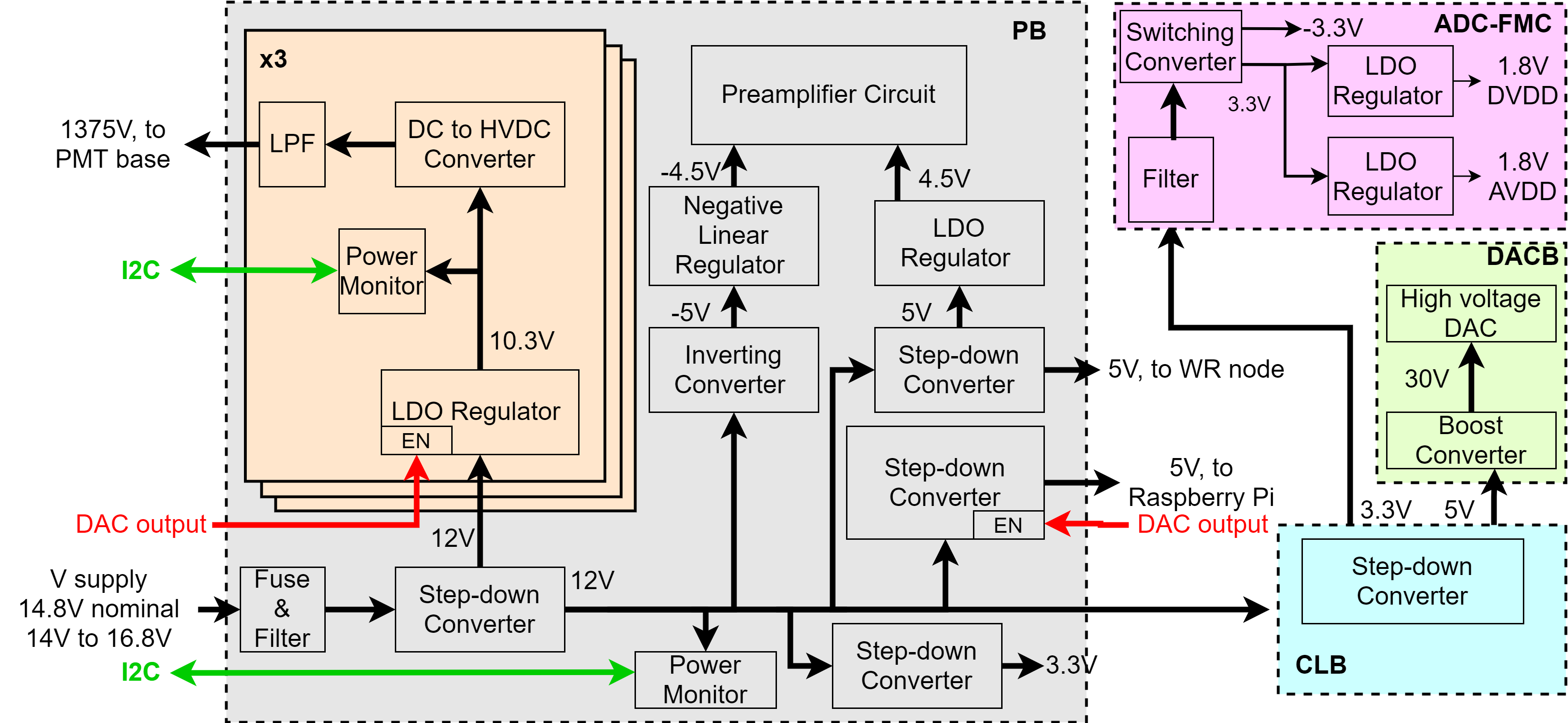}}
	\caption{Block diagram of the power supply in modules. The power board regulates the main power input and provides power to the whole module. The power supply of the Raspberry Pi, PMTs, and LEDs is controlled remotely by the DAQ. The power monitoring circuits are connected to significant nodes.}
	\label{fig:powers}
\end{figure}

As shown in \figurename~\ref{fig:e_o_d}, a lithium battery in the control and battery module supplies power to the LEM and LRMs through electro-optical cables.
Each module has a number of electronic boards including commercial boards (WR node, Raspberry Pi, and CLB) and custom-designed hardware (ADCB, digital-to-analog mezzanine board, PB, and LED board). The power supply of the Raspberry Pi, PMTs, and LEDs is controlled remotely by the DAQ. At power-up, all of them are pulled down to the ground to prevent uncontrolled turn-on. The schematic view of the power supply in modules is illustrated in \figurename~\ref{fig:powers}.

The PB, shown in \figurename~\ref{fig:pb}, regulates the main power input (14 V to 16.8 V) and provides power to the whole module. The battery-powered input first passes through a Schottky diode to prevent reverse currents. The power monitoring circuits are connected to significant nodes. In addition, analog circuits use the low-dropout linear regulator (LDO) as the power supply to reduce power supply noise. A high voltage module (XP Power GP15) converts 10.3 V to 1375 V with a low-pass filter so that the ripple and noise are reduced to below 4 mVpp. The power consumption of high voltage generation of each PMT is approximately 400 mW.

Two high voltage DACs are designed in the digital-to-analog mezzanine board (DACB) (see \figurename~\ref{fig:CLBbs}(c)). In the LRM, only one DAC is used to control the high voltage of the three PMTs and the power supply of the Raspberry Pi. Each DAC in the LEM correspondingly controls four LED driver circuits on an LED board (LEDB), enabling LED drivers (ADI LT3465) or providing a reference level for the pulsed LED driver circuit \cite{Lubsandorzhiev:2004zh,Kap}. The DAC can provide a maximum voltage of 30 V with a minimum adjustable step of 7 mV to the pulsed LED driver circuit so that it emits light pulses of sufficient intensity. In the global trigger mode, the brightness of the pulsing LED is adjusted such that the dominant signals in PMT outputs are SPE signals. A detailed introduction and test results of LEDB are described in \cite{TRIDENTLED}.

\subsection{Central logic board Firmware}
\label{subsec:firmware}
\begin{figure}[htbp]
	\centerline{\includegraphics[width=3.5in]{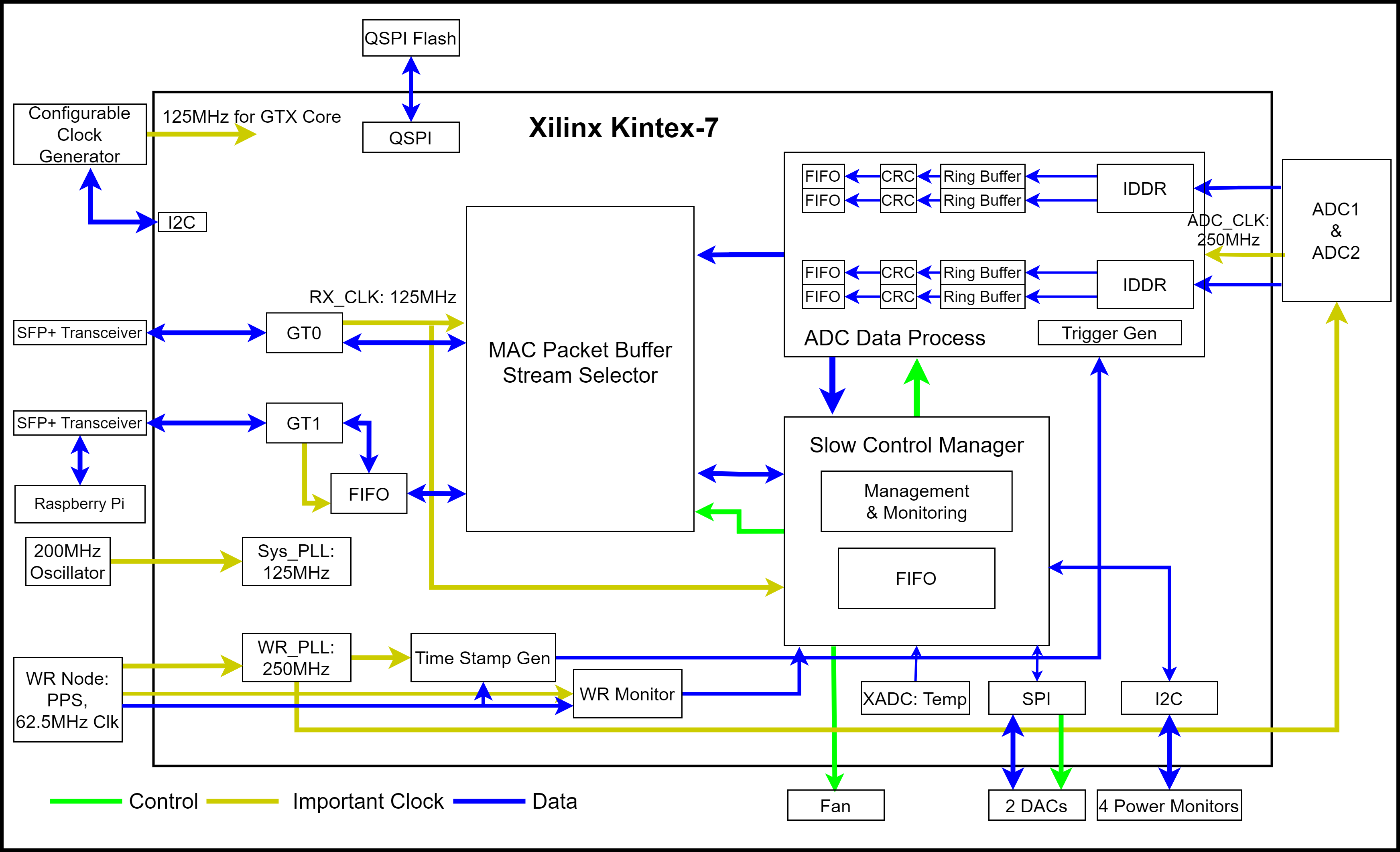}}
	\caption{Block diagram of the CLB FPGA firmware.}
	\label{fig:adcfirmware}
\end{figure}

The FPGA firmwares running in both LRMs and LEMs are the same. The block diagram of the control and readout logic is shown in \figurename~\ref{fig:adcfirmware}. Its main blocks are ADC data process, slow control manager, and MAC packet manager, which are described as following. 

\subsubsection{ADC data process}
\begin{figure}[htbp]
	\centerline{\includegraphics[width=3.5in]{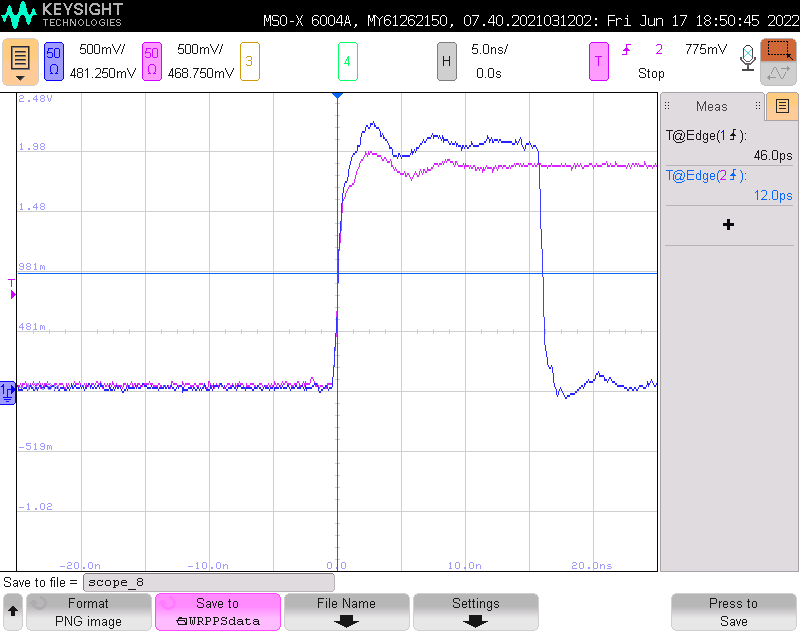}}
	\caption{The trigger signal of the module (blue line) and the PPS of the WR switch (pink line) are displayed on the Keysight MSO-X 6004A oscilloscope. This demonstrates triggers are synchronized after fiber link calibration.}
	\label{fig:tralign}
\end{figure}

As mentioned in Section~\ref{subsec:PMT}, the ADCs digitize PMT signals with a sampling rate of 250 MSPS. Each ADC outputs a 250 MHz clock and data, the latter conveys two channels of ADC data which are decoupled via input double-data-rate (IDDR) registers. The firmware supports two trigger modes: self-trigger mode and global trigger mode. In the self-trigger mode, signals exceeding the threshold are captured with corresponding time stamps. In the global trigger mode, all CLBs generate trigger signals synchronously at the same frequency, which is aligned with the WR clock. This trigger signal causes the LED in the LEM to pulse and events to be generated in the LRM. The measurement result of the trigger signal of the module is shown in \figurename~\ref{fig:tralign}, in which the trigger signal is aligned with the PPS of the WR switch precisely after calibrating the uplink and downlink delays of the optical fiber links.

The real-time data stream is buffered in a ring buffer. An event includes a segment of data with the predetermined window length in the ring buffer, a channel number, a timestamp, and an event number. Some parameters are adjusted by the slow control such as the length of the event window (the maximum of 500 sampling points), the position of the trigger, the threshold of the trigger, and the mode of the trigger generator. And then a cyclic redundancy check (CRC) is added to the end of the event packet before being transmitted to an asynchronous FIFO. 

\subsubsection{Slow control manager}
The slow control manager handles various affairs, including executing commands sent by the host-computer, monitoring the system, and uploading feedback and running states. The commands are decoded from the received media access control (MAC) packets. The result of the command execution is updated into the corresponding slow control register. The slow control registers comprise two types: read-only or writable registers. The read-only registers store status monitoring and sensor data. For example, the status of each channel whether the ring buffers or the FIFO overflow during data acquisition are updated into read-only registers in real time. The values in the writable registers determine the configuration of data acquisitions and DAC outputs. Notably, after each configuration of the writable register, it will be read back immediately to verify whether the configuration is successful.

\subsubsection{MAC packet manager}
The data transmission between the CLB and the switch board complies with the MAC protocol, which has a better payload and higher throughput than TCP and UDP. The network bandwidth is over 400 Mbps, which supports a total trigger rate of 40 kHz per LRM with a 1-microsecond sample window. The MAC packet contains echo data (for debugging), heartbeat frames, slow control data, ADC data, and Raspberry Pi data, which are obtained from the corresponding FIFO. 

The sending and receiving of MAC packets are realized by 1.25 Gbps Xilinx GTX IP core \cite{gtx}, the communication with the switch board runs on the GT0 channel, and the communication with the Raspberry Pi runs on the GT1 channel. Once the MAC packet comes from the Raspberry Pi, it will be retransmitted to the GT0 channel by the MAC packet manager. 

\subsection{Data acquisition system}
\label{subsec:host-com}
The readout of the T-REX detector is based on the all-data-to-vessel concept in which all digitized signals from the PMTs are sent to DAQ for real-time processing. The DAQ comprises a stand-alone WR switch, a switch board, and a host-computer. 

The switch board is an intermediate station between CLBs in modules and the host-computer to distribute commands and aggregate data. Its hardware is the CLB (see \figurename~\ref{fig:CLBbs}(a)) using four SFPs, which are respectively connected to the host-computer, the first LRM, the second LRM, and the LEM. The firmware of the switch board comprises a GTX IP core and a MAC packet manager that are similar to the description in Section~\ref{subsec:firmware}.

\begin{figure}[htbp]
	\centerline{\includegraphics[width=3.5in]{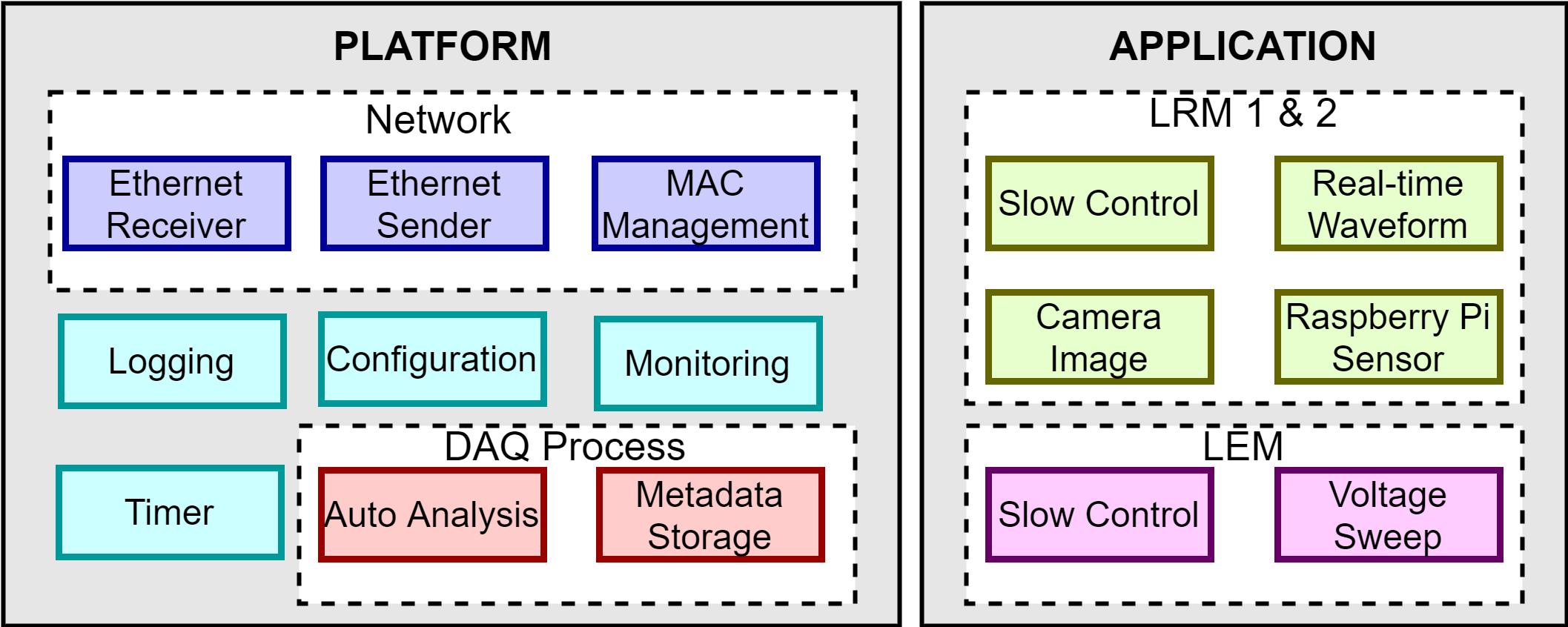}}
	\caption{Representation of the layers and structure of the host-computer software.}
	\label{fig:software}
\end{figure}

The host-computer software is the brain of the electronics system, responsible for sending commands and receiving data. The software running on the Windows system is coded in C++ with QT5. As illustrated in \figurename~\ref{fig:software}, the software is structured in the platform layer and application layer. Each block of functions occupies a separate thread.

The platform layer contains functions for the network, the DAQ process, and common use. MAC frames of the specified format are sent and received through MAC management, Ethernet sender, and Ethernet receiver. The decoded ADC data, Raspberry Pi data, and slow control data are transferred to the corresponding buffers for further processing. Auto analysis parses the data in each buffer to form waveforms, camera images, WR time, and slow control register values. All data can be saved in binary format through metadata storage for offline analysis. Common functions are used throughout the software. For example, the experiment's configuration can be saved as a JSON file, and the previous configuration file can be loaded with authority.

\begin{figure}[htbp]
	
	\centerline{\includegraphics[width=3.5in]{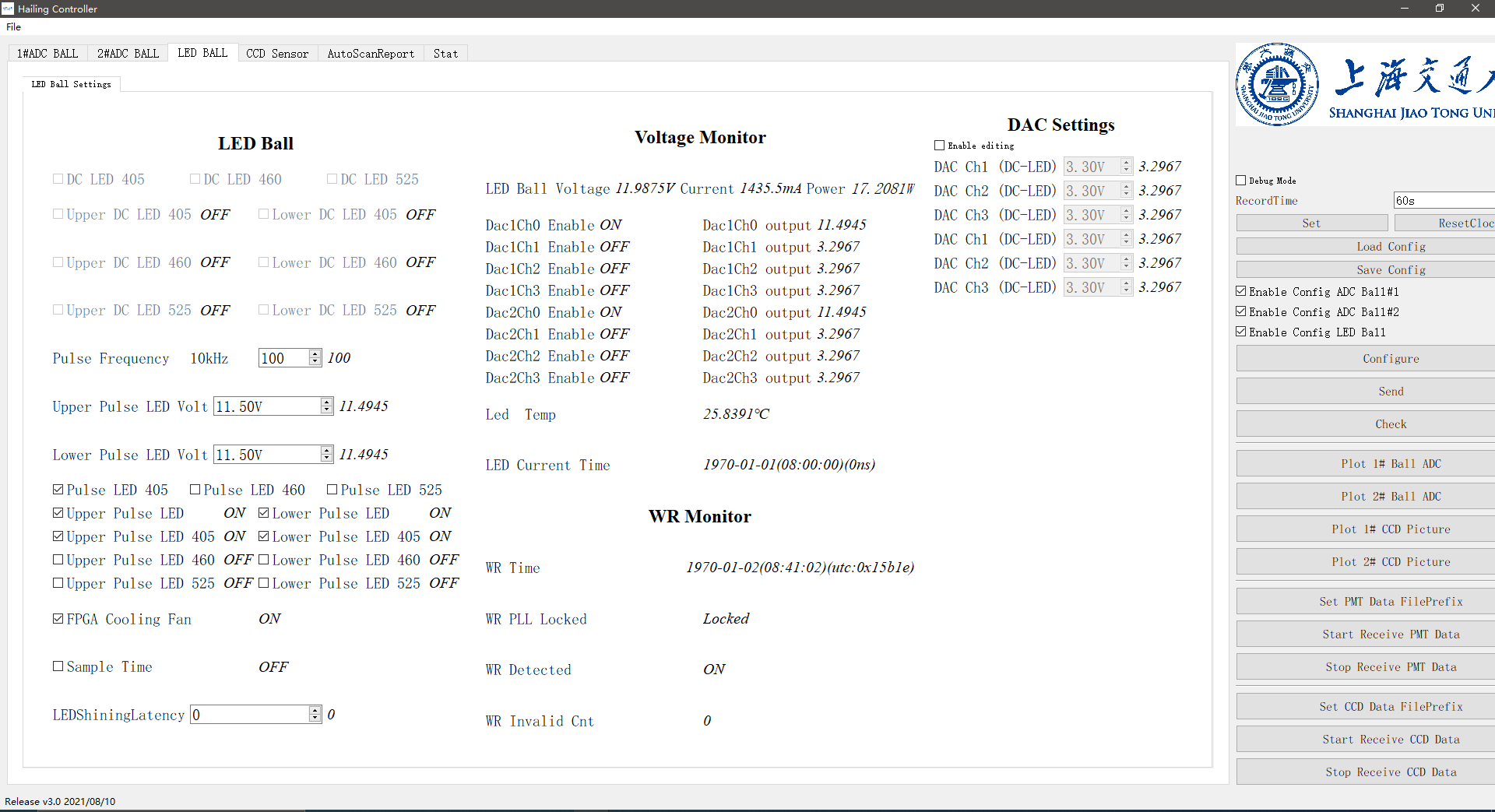}}
	\centerline{\includegraphics[width=3.5in]{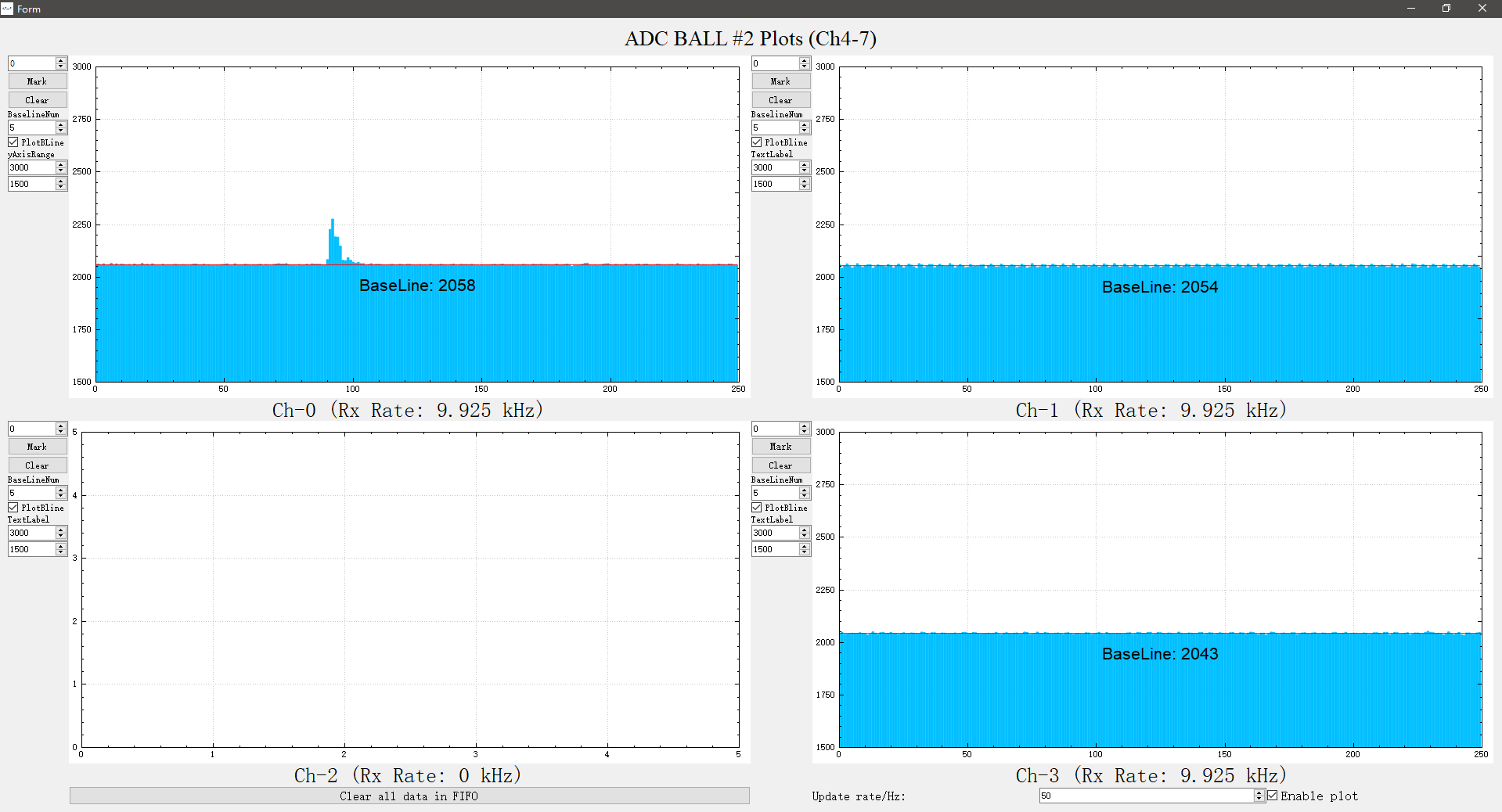}}
	\caption{Screenshots of the user interface of the host-computer. Top: the user interface of the slow control. Bottom: real-time data analysis and refreshed waveform plotting.}
	\label{fig:ui}
\end{figure}

The application layer includes high-level functionalities provided to users. The slow control, in the main thread, is responsible for configuring the slow control registers of each module inner CLB, and updating the user interfaces (UIs) after reading back the values of these registers, as shown in the upper panel of \figurename~\ref{fig:ui}. Images recorded by the camera and other data from sensors are displayed in separate windows. The lower panel of \figurename~\ref{fig:ui} shows the real-time waveforms and supports baseline calculation and event rate estimation. The function of voltage sweep is to gradually change the LED brightness within the set range and automatically analyze the ADC data to obtain the ratio of the SPE signal.

\section{Results of laboratory commissioning}
\label{sec:data}
Electronic commissioning with photosensitive devices was conducted in the laboratory before deploying at sea to validate the functionality and stability of the readout electronics described above.

\begin{figure}[htbp]
	\centerline{\includegraphics[width=3.5in]{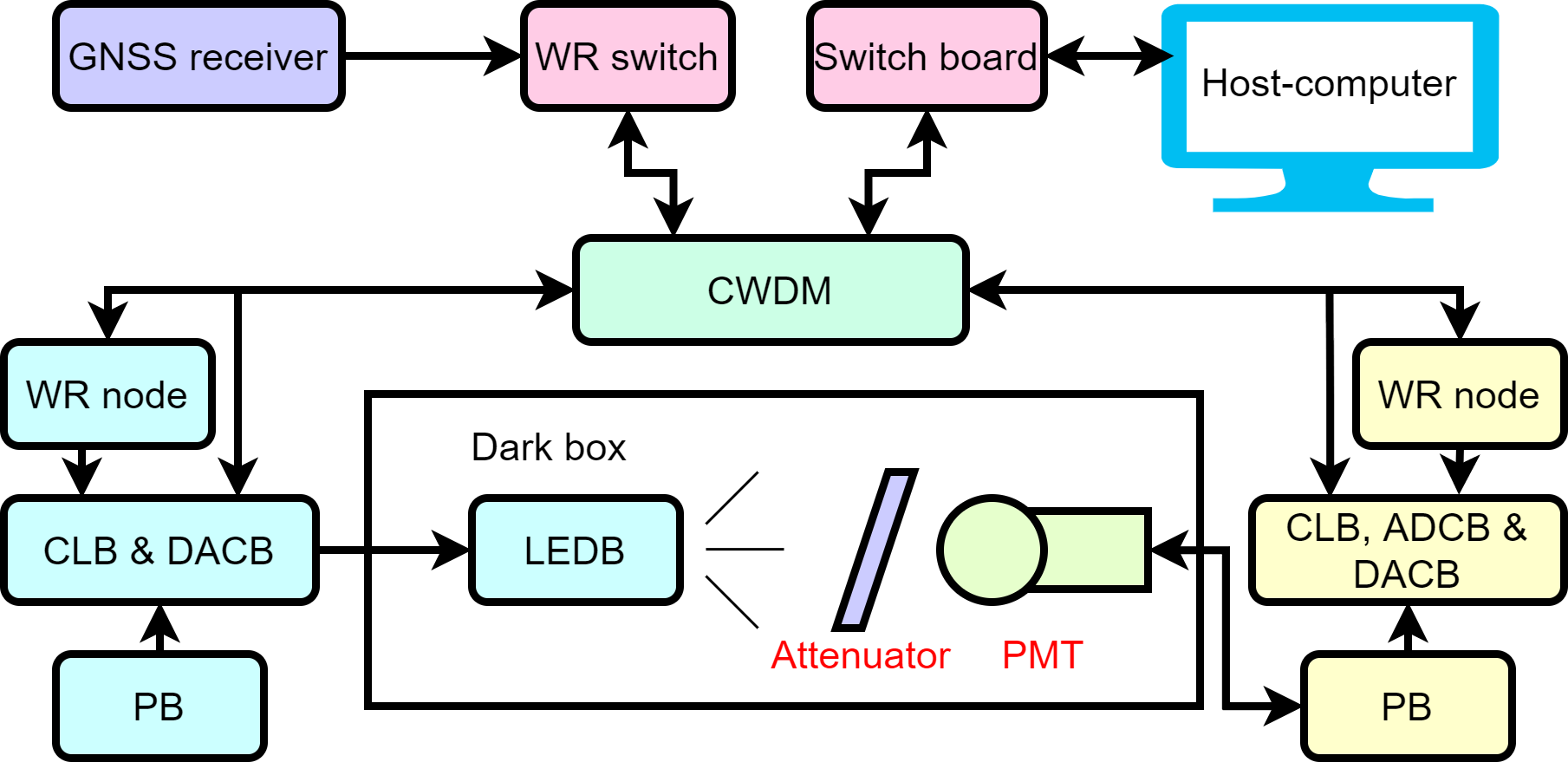}}
	\caption{Sketch of the test bench for PMT calibration. The WR system is combined with the global navigation satellite system (GNSS) receiver.}
	\label{fig:labtest}
\end{figure}

\begin{figure}[htbp]
	\centerline{\includegraphics[width=3.5in]{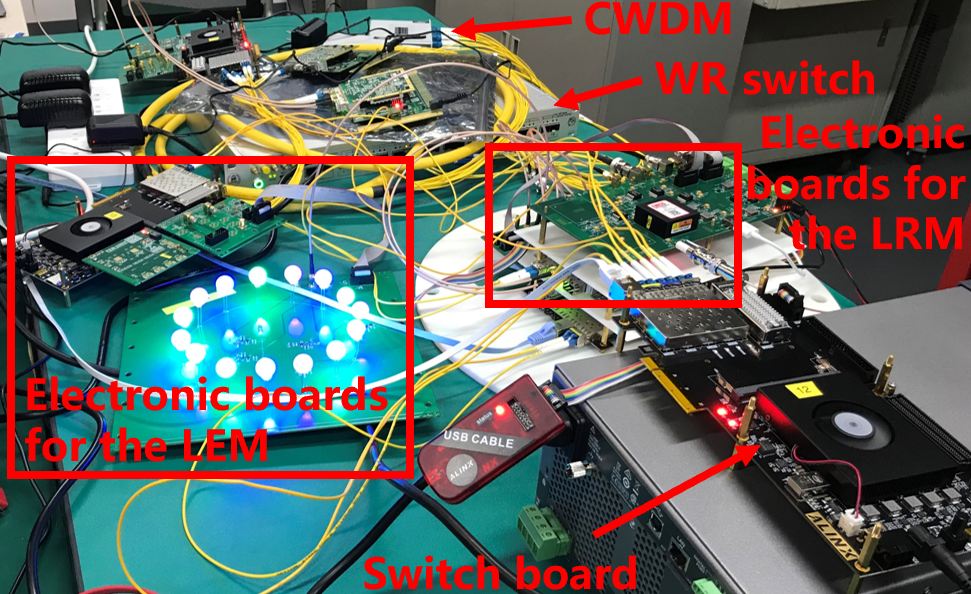}}
	\caption{The setup of the test bench for commissioning before module assembly. The dark box with PMTs and LEDBs is placed under the anti-static table.}
	\label{fig:labtestsetup}
\end{figure}

The electrical performance of all 3-inch PMTs was measured in the laboratory \cite{TRIDENTPMT}. The test bench displayed in \figurename~\ref{fig:labtest} uses the whole readout electronics. An LEDB and a PMT are placed in the dark box. An optical attenuator is inserted in front of the PMT to simulate the light attenuation over tens of meters in seawater.  \figurename~\ref{fig:labtestsetup} shows the setup for commissioning before module assembly. A total of six PMTs are evaluated simultaneously. In this configuration, the maximum trigger rate that each channel can handle is up to 30 kHz.

\begin{figure}[htbp]
	\centerline{\includegraphics[width=3.5in]{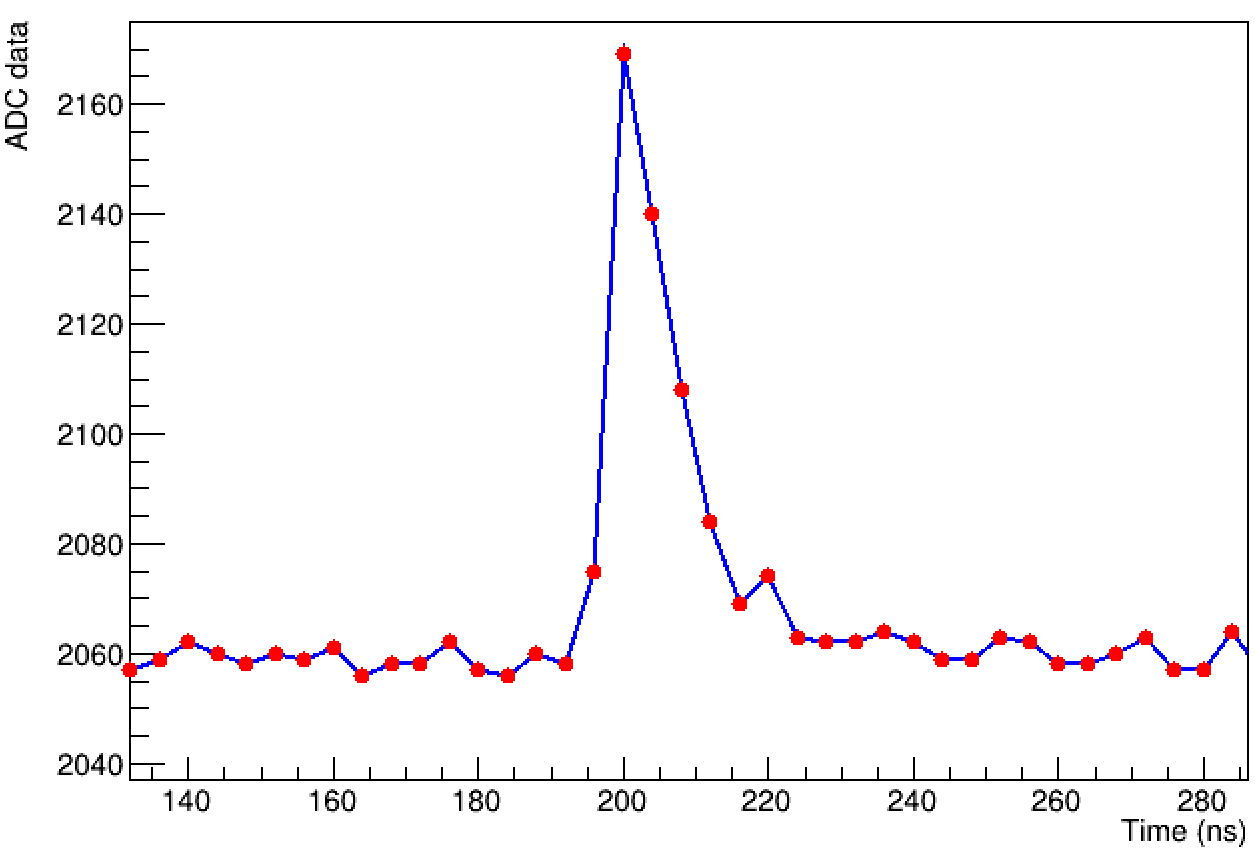}}
	\caption{An example of the PE waveform from one measured PMT.}
	\label{fig:SPEdata1}
\end{figure}

\begin{figure}[htbp]
	\centerline{\includegraphics[width=3.5in]{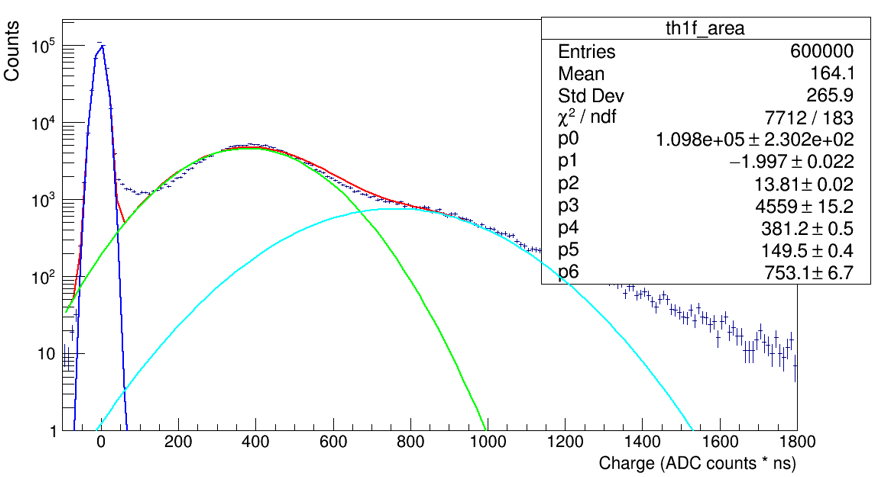}}
	\caption{A typical SPE spectrum for one XP72B22 in units of ADC counts times nanoseconds. p0 to p6 are fitting parameters obtained by three-Gaussian fitting.}
	\label{fig:SPEdata2}
\end{figure}

The typical digitized PMT signal is shown in \figurename~\ref{fig:SPEdata1} and the typical SPE spectrum for one XP72B22 is illustrated in \figurename~\ref{fig:SPEdata2}, in which the pedestal, single PE peak, and double PE peak can be identified. The spectrum is fitted with a three-Gaussian function:
\begin{equation}
	%\label{eq1}
	\begin{split}
		f(Q)=p_{0}e^{-\frac{(Q-p_{1})^{2}}{2p^{2}_{2}}}+p_{3}e^{-\frac{(Q-p_{1}-p_{4})^{2}}{2(p^{2}_{2}+p^{2}_{5})}} \\ +p_{6}e^{-\frac{(Q-p_{1}-2p_{4})^{2}}{2(p^{2}_{2}+2p^{2}_{5})}} .
	\end{split}	
\end{equation}
The PMT's gain is calculated by converting the single PE peak's position.

\begin{figure}[htbp]
	%\centerline{\includegraphics[width=2.7in]{timeoffset_lab.png}}
	\centerline{\includegraphics[width=3.5in]{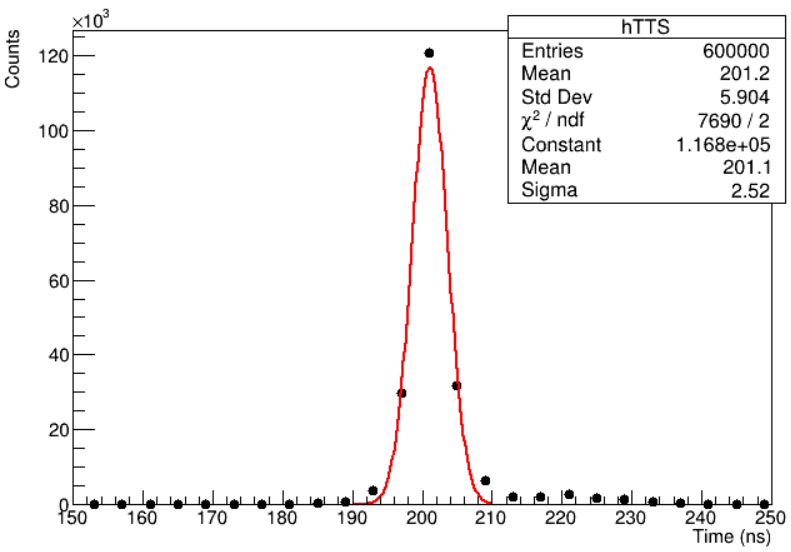}}
	\caption{An example of arrival time measurement results. The spread of arrival time comprises electronic jitter, the uncertainty of luminescence, and the transit time spread of the PMT.}
	\label{fig:tts}
\end{figure}

The fast time response characteristic of LED light pulse emission is an important factor in the experiment. The same setup can be used to evaluate LEDs from different vendors. The arrival time when the photon emitted by the pulsing LED reaches the PMT is measured. An example of arrival time distribution is shown in \figurename~\ref{fig:tts}. The smaller the spread of the arrival time, the better the time characteristic of the LED.

\begin{figure}[htbp]
	\centerline{\includegraphics[width=3.5in]{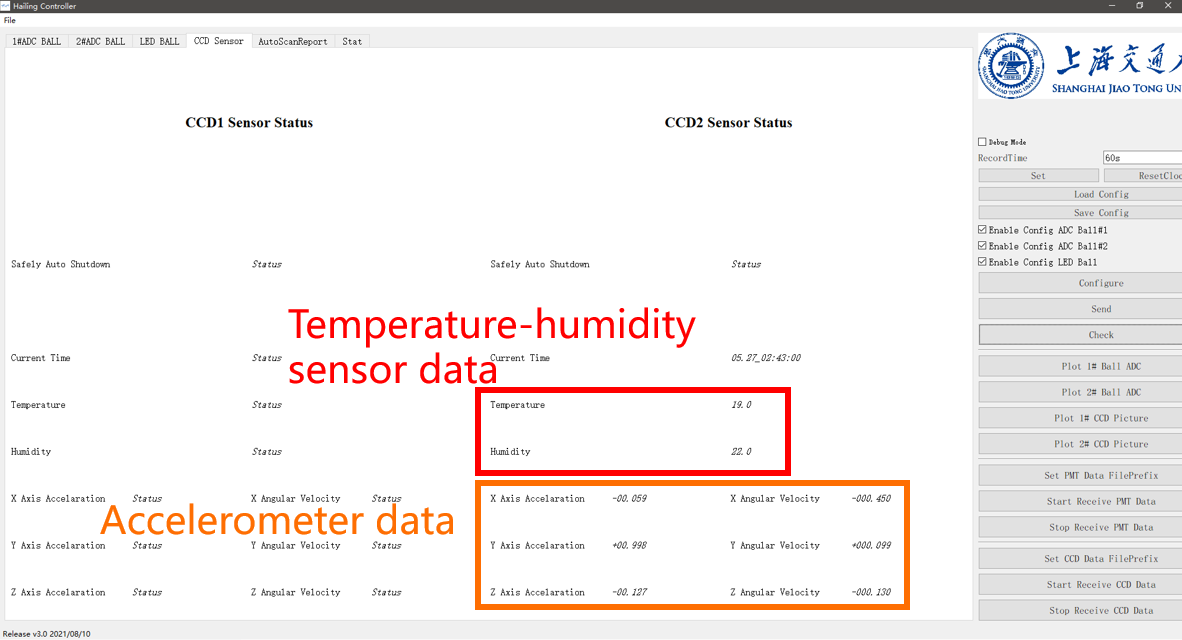}}
	\centerline{\includegraphics[width=3.5in]{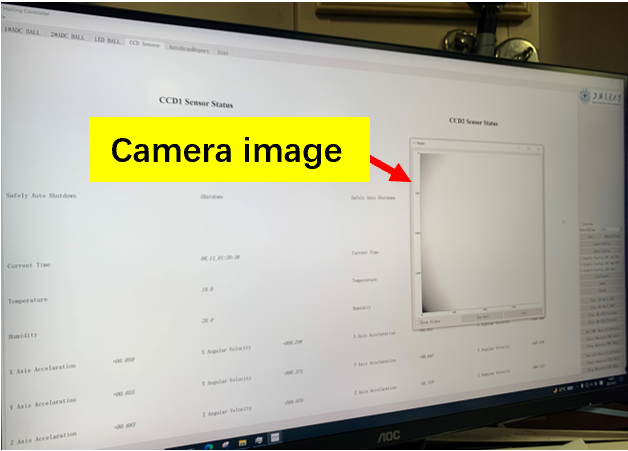}}
	\caption{Top: the data from the temperature-humidity sensor and accelerometer are transferred to the host computer. Bottom: images captured by the camera system are transferred to DAQ in real time and displayed.}
	\label{fig:ccdui}
\end{figure}

The host-computer analyzes and displays sensor (temperature-humidity sensor and accelerometer) data in real time, shown in the upper panel of \figurename~\ref{fig:ccdui}. The measured temperature and humidity are used to monitor the state inside the module. The bottom panel of \figurename~\ref{fig:ccdui} shows an image captured by the camera during the experiment. These images can be used to aid in the deployment of the T-REX apparatus.

\section{Conclusion}
\label{sec:conclusions}
In this paper, the readout electronics of the T-REX has been presented. The main electronic boards, FPGA firmware, and host-computer software are described in detail. The full chain of the readout electronics is tested thoroughly in the laboratory for its functionality and performance. The readout system developed above fulfilled all requirements for its application and proved to be robust \emph{in-situ} during the T-REX mission to the South China Sea in September 2021 at a depth of 3420 m.

\appendices

\section*{Acknowledgment}
The authors would like to thank Jun Guo, Fan Hu, Wenlian Li, Iwan Morton-Blake, Wei Tian, and Fuyudi Zhang for their help to improve this paper.

\end{document}